**Remaining problems with the "New Crown Indicator" (MNCS) of the CWTS**
*Letter to the Editor*

Loet Leydesdorff [1] & Tobias Opthof [2]

In their article, entitled "Towards a new crown indicator: some theoretical considerations," Waltman *et al*. (2010) show that the "old crown indicator" of CWTS in Leiden was mathematically inconsistent and that one should move to the normalization as applied in the "new crown indicator." Although we now agree about the statistical normalization (Opthof & Leydesdorff, 2010; Van Raan *et al*., 2010), the "new crown indicator" inherits the scientometric problems of the "old" one in treating subject categories as a standard for normalizing differences in citation behavior among fields of science. If a paper is published in a journal with more than one attributed ISI Subject Category, the authors propose that the paper should equally be weighted as belonging for a percentage to these different fields. Instead of averaging these fractions, they favour the harmonic mean because the property of mathematical consistency is then maintained in the construct.

Because the Leiden indicators can be expected to remain important in scientometric evaluation studies, we would like to mention the following problems with this field normalization:

1. The ISI Subject Categories were not designed for the scientometric evaluation, but for the purpose of information retrieval. The subject categories lack an analytical base (Boyack *et al*., 2005; Pudovkin & Garfield, 2002, at p. 1113n.; Rafols & Leydesdorff, 2009) and are not regularly updated with the development of the scientific literature (Bensman & Leydesdorff, 2009). Alternative and far more precise classification schemes are nowadays available. Why not evaluate an academic hospital on the basis of the Medical Subject Headings (MeSH) of the bibliographic database MedLine, which are publicly available and compiled on a paper-by-paper basis (Bornmann *et al*., 2008, at p. 98)?

2. If papers are published in journals which are attributed to several subject categories, CWTS chooses to weigh each category equally. This procedure generates artifacts in the rankings, since some journals are highly specialized (e.g., in cardiology) but nevertheless subsumed under a number of categories (for the purpose of information retrieval). Without casting doubt on the primary category, the attribution to more than a single category can be functional to open the information to additional fields. The attribution of categories to journals, however, is not based on multivariate analysis of the citation matrix among journals or weighted in terms of numbers of citations (Leydesdorff, 2006; Pudovkin & Garfield, 2002).

[1] Amsterdam School of Communications Research (ASCoR), University of Amsterdam, Kloveniersburgwal 48, 1012 CX Amsterdam, The Netherlands; loet@leydesdorff.net.
[2] Experimental Cardiology Group, Heart Failure Research Center, Academic Medical Center AMC, Meibergdreef 9, 1105 AZ Amsterdam, The Netherlands; t.opthof@inter.nl.net.

For example, the *Journal of Vascular Research* is attributed to the subject categories of "peripheral vascular disease" and "physiology," and the journal *Circulation* to "cardiac and cardiovascular systems," "hematology," and "peripheral vascular diseases"; whereas the *American Journal of Cardiology* is attributed only to "cardiac and cardiovascular systems." Scholars in these fields, however, publish and cite across such categorical divides. Use of these subject categories thus generates (uncontrolled) indexer effects.

3. The purpose of normalization at the field level is to control for differences in "citation potentials" among fields (Garfield, 1979). These differences are caused by differences in citation behavior among scholars in various fields of science. Mathematics, for example, is known to have a much lower citation density than the biomedical sciences (McAllister *et al*., 1983). For example, if an author in mathematics cites six references, each reference can be counted as 1/6 of overall citation, whereas a citation in a paper in biomedicine with 40 cited references can be counted as 1/40. This normalization thoroughly takes field differences into account and the results allow for statistical testing (Leydesdorff & Bornmann, in press).

   Most importantly, fractional counting is independent from a classification system and thus there is no indexer effect. It is also independent of journals as implicit frames of reference for the ISI Subject Categories. Different topics in the same journal may be very differently cited. Although Waltman *et al*. (2010) acknowledge the possibility to capture the differences in citation behavior by fractional counting in the citing articles (Leydesdorff & Opthof, 2010a and 2010b; Moed, 2010; Zitt & Small, 2008), the "new crown indicator" remains consistent with the "old" one by choosing for field-normalization in terms of the cited documents organized in journals.

Since we seem presently at the brink of developing a new "crown indicator," we further wish to suggest that these authors take into account that the "mean" is not a proper statistics for measuring differences among skewed distributions. Without changing the acronym of "MNCS," one could define the "Median Normalized Citation Score." This would relate the new crown indicator directly to the percentile approach that is, for example, used in the *Science and Engineering Indicators* of US National Science Board (2010; cf. Garfield & Pudovkin, in press; Plomp, 1990). The median is by definition equal to the 50$^{th}$ percentile. The indicator can thus easily be extended with the 1% (= 99$^{th}$ percentile) most highly-cited papers (Bornmann *et al*., in press). The seeming disadvantage of having to use non-parametric statistics is more than compensated by possible gains in the precision.

**References**:

Bensman, S. J., & Leydesdorff, L. (2009). Definition and Identification of Journals as Bibliographic and Subject Entities: Librarianship vs. ISI Journal Citation Reports (JCR) Methods and their Effect on Citation Measures. *Journal of the American Society for Information Science and Technology, 60*(6), 1097-1117.

Bornmann, L., Mutz, R., Neuhaus, C., & Daniel, H. D. (2008). Citation counts for research evaluation: standards of good practice for analyzing bibliometric data and presenting and interpreting results. *Ethics in Science and Environmental Politics(ESEP), 8*(1), 93-102.

Bornmann, L., de Moya Anegón, F., & Leydesdorff, L. (in press). Does scientific advancement lean on the shoulders of mediocre research? An investigation of the Ortega hypothesis. *PLoS ONE*.

Boyack, K. W., Klavans, R., & Börner, K. (2005). Mapping the Backbone of Science. *Scientometrics, 64*(3), 351-374.

Garfield, E. (1979). Is citation analysis a legitimate evaluation tool? *Scientometrics, 1*(4), 359-375.

Leydesdorff, L. (2006). Can Scientific Journals be Classified in Terms of Aggregated Journal-Journal Citation Relations using the Journal Citation Reports? *Journal of the American Society for Information Science & Technology, 57*(5), 601-613.

Leydesdorff, L., & Bornmann, L. (2011). How fractional counting affects the Impact Factor: Normalization in terms of differences in citation potentials among fields of science. *Journal of the American Society for Information Science and Technology, in press*.

Leydesdorff, L., & Opthof, T. (2010a). Normalization at the field level: fractional counting of citations. *Journal of Informetrics, 4*(4), 644-646.

Leydesdorff, L., & Opthof, T. (2010b). *Scopus*' Source Normalized Impact per Paper (SNIP) *versus* the Journal Impact Factor based on fractional counting of citations. *Journal of the American Society for Information Science and Technology, DOI: 10.1002/asi.21371*.

McAllister, P. R., Narin, F., & Corrigan, J. G. (1983). Programmatic Evaluation and Comparison Based on Standardized Citation Scores. *IEEE Transactions on Engineering Management, 30*(4), 205-211.

Moed, H. (2010). CWTS crown indicator measures citation impact of a research group's publication oeuvre. *Journal of Informetrics, 3*(3), 436-438.

National Science Board. (2010). *Science and Engineering Indicators*. Washington DC: National Science Foundation; available at http://www.nsf.gov/statistics/seind10/.

Opthof, T., & Leydesdorff, L. (2010). *Caveats* for the journal and field normalizations in the CWTS ("Leiden") evaluations of research performance. *Journal of Informetrics, 4*(3), 423-430.

Plomp, R. (1990). The significance of the number of highly cited papers as an indicator of scientific prolificacy. *Scientometrics, 19*(3), 185-197.

Pudovkin, A. I., & Garfield, E. (2002). Algorithmic procedure for finding semantically related journals. *Journal of the American Society for Information Science and Technology, 53*(13), 1113-1119.

Pudovkin, A. I., & Garfield, E. (in press). Percentile rank and author superiority indexes for evaluating individual journal articles and the author's overall citation performance. *CollNet Journal*.

Rafols, I., & Leydesdorff, L. (2009). Content-based and Algorithmic Classifications of Journals: Perspectives on the Dynamics of Scientific Communication and Indexer Effects *Journal of the American Society for Information Science and Technology, 60*(9), 1823-1835.

Van Raan, A. F. J., Van Leeuwen, T. N., Visser, M. S., van Eck, N. J., & Waltman, L. (2010). Rivals for the crown: Reply to Opthof and Leydesdorff. *Journal of Informetrics, 4*(3), 431-435.

Waltman, L., Van Eck, N. J., Van Leeuwen, T. N., Visser, M. S., & Van Raan, A. F. J. (2010). Towards a new crown indicator:Some theoretical considerations. *Journal of Informetrics*, doi:10.1016/j.joi.2010.08.001.

Zitt, M., & Small, H. (2008). Modifying the journal impact factor by fractional citation weighting: The audience factor. *Journal of the American Society for Information Science and Technology, 59*(11), 1856-1860.